\documentclass[12pt]{article}
\usepackage{feynmf}
\usepackage{graphicx}
\usepackage{epsfig}
\newcommand{\beq}{\begin{equation}}
\newcommand{\eeq}{\end{equation}}
\newcommand{\beqn}{\begin{eqnarray}}
\newcommand{\eeqn}{\end{eqnarray}}
\begin{document} 
 
\title{About estimations of an one-photon background in neutrino oscillation
experiments at low neutrino energies} 
\author{V.P. Efrosinin\\
Institute for Nuclear Research RAS\\
60th October Ave., 7A, 117312, Moscow, Russia}

\date{}
\renewcommand {\baselinestretch} {1.3}

\maketitle
\begin{abstract}

The possible sources of one-photon radiation as background for quasi-elastic
reaction $\nu_{\mu}+n \to \mu^-+p$ are considered. The ones are relevant in
experiments on determination of oscillation parameters at low neutrino energies
($E_{\nu} \sim 1~GeV$). The estimation for the cross section of reaction
$\nu_{\mu}+n \to \mu^-+p+\gamma$ is given at $E^{lab}_{\nu} = 0.7~GeV$ as
$0.65\%$ from the corresponding cross section of quasi-elastic reaction.
The mechanisms of quasi-elastic reaction are also considered at low neutrino
energies on quark level.    
\end{abstract}

The current now accelerator experiments on $\nu_{\mu}-\nu_e$ neutrino
oscillation deal with neutrino beams at low energies ($E_{\nu} \sim 1~GeV$).
To shed liht on the neutrino oscillation parameters,
the rather precise knowledge of the neutrino-nucleon interactions in this
energy region is required. Since Cherenkov detectors employed in these
experiments poorly distinguish an electron from a photon, it is extremely
important to study the possible background from other reactions with one or two
photons to the main quasi-elastic reaction a neutrino on a nucleon:
\begin{eqnarray}
\label{eq:MM1}
\nu_{\mu}+n \to \mu^-+p.  
\end{eqnarray}

In this work we give estimations of the cross-sections of the processes,
wich are interested for us, at neutrino
energy in laboratory system $E^{lab}_{\nu} = 0.7~GeV$.

The processes with neutral currents
\begin{eqnarray}
\label{eq:MM2}
\nu_{\mu}+p \to \nu_{\mu}+p,\nonumber\\
\nu_{\mu}+n \to \nu_{\mu}+n  
\end{eqnarray}
are not interested in experimental analysis of neutrino oscillations.
However, reaction
\begin{eqnarray}
\label{eq:MM3}
\nu_{\mu}+p \to \nu_{\mu}+p+\gamma  
\end{eqnarray}
can contribute to the one-photon background. At the same time,
the ratio of the cross sections in the given neutrino
energy is known from experiments \cite{Zeller}:
\begin{eqnarray}
\label{eq:MM4}
(\nu_{\mu}p\to\nu_{\mu}p)/(\nu_{\mu}n\to\mu^-p) \sim 0.1.  
\end{eqnarray}
The $Q^2$ interval, over which this ratio was measured, is $(0.3\div1.0)~GeV^2$.
The approximately same ratio is valid for the cross sections of the one-photon
reactions
\begin{eqnarray*}
(\nu_{\mu}p\to\nu_{\mu}p\gamma)/(\nu_{\mu}n\to\mu^-p\gamma).  
\end{eqnarray*}
The following contribution of reaction (\ref{eq:MM3}) is essentially less
than the contribution of reaction
\begin{eqnarray}
\label{eq:MM5}
\nu_{\mu}+n \to \mu^-+p+\gamma.  
\end{eqnarray}
The calculation of the cross section of reaction (\ref{eq:MM5}) is the first
goal of the present study.

The two-photon background in our case is determined by reaction
\begin{eqnarray}
\label{eq:MM6}
\nu_{\mu}+n \to \mu^-+p+\pi^0.  
\end{eqnarray}
The experimental cross section of reaction (\ref{eq:MM6}) at
$E^{lab}_{\nu} = 0.7~GeV$ is $\sigma \simeq 0.07 \times 10^{-38}~cm^2$
\cite{Zeller}. The pair of photons from $\pi^0$ decay is successfully
identified by data processing of experiments except of the
restricted kinematic configurations.

Reaction (\ref{eq:MM6}) together with reaction
\begin{eqnarray}
\label{eq:MM7}
\nu_{\mu}+n \to \mu^-+n+\pi^+  
\end{eqnarray}
can also contribute to the one-photon background. The experimental
cross section of reaction
(\ref{eq:MM7}) is approximately the same as that of reaction (\ref{eq:MM6})
 and at $E^{lab}_{\nu} = 0.7~GeV$ it equals to
$\sigma \simeq 0.07 \times 10^{-38}~cm^2$. If the isobar $\Delta^+$ is
produced in the reactions (\ref{eq:MM6}) and (\ref{eq:MM7}), it can decay
via the channel:
\begin{eqnarray}
\label{eq:MM8}
\Delta^+ \to p+\gamma.  
\end{eqnarray}
The branching ratio of the decay (\ref{eq:MM8}) is $\sim 0.5\%$
\cite{part}. We are suspected that reactions (\ref{eq:MM6}) and
(\ref{eq:MM7}) go through the formation of an isobar $\Delta^+$. The common
contribution to the one-photon background from the reactions
(\ref{eq:MM6},\ref{eq:MM7}) as well as from the neutral current single
pion production
$\nu_{\mu}p \to \nu_{\mu}p\pi^0$ is
\begin{eqnarray}
\label{eq:MM9}
\sigma \le (0.07+0.07+0.04)*0.005 \times 10^{-38}~cm^2 \simeq
0.0009 \times 10^{-38}~cm^2.  
\end{eqnarray}
It constitutes $< 0.09\%$ from the cross section of quasi-elastic reaction.
The actual magnitude of one-photon cross section (\ref{eq:MM9})
will be essentialy less that indicated value,
because considerable non-resonant pion-nucleon contribution there is in the
final state.

Therefore, further we will consider only of reactions
(\ref{eq:MM1}) and (\ref{eq:MM5}). It should be noted that
without comprehension of the mechanism of
quasi-elastic reaction, it is not possible to write out the gauge-invariant
amplitude
of reaction (\ref{eq:MM5}), because this reaction (\ref{eq:MM5}) is
radiative process from the process (\ref{eq:MM1}).

\begin{figure*}[hb]
\begin{center}
\epsfig{file=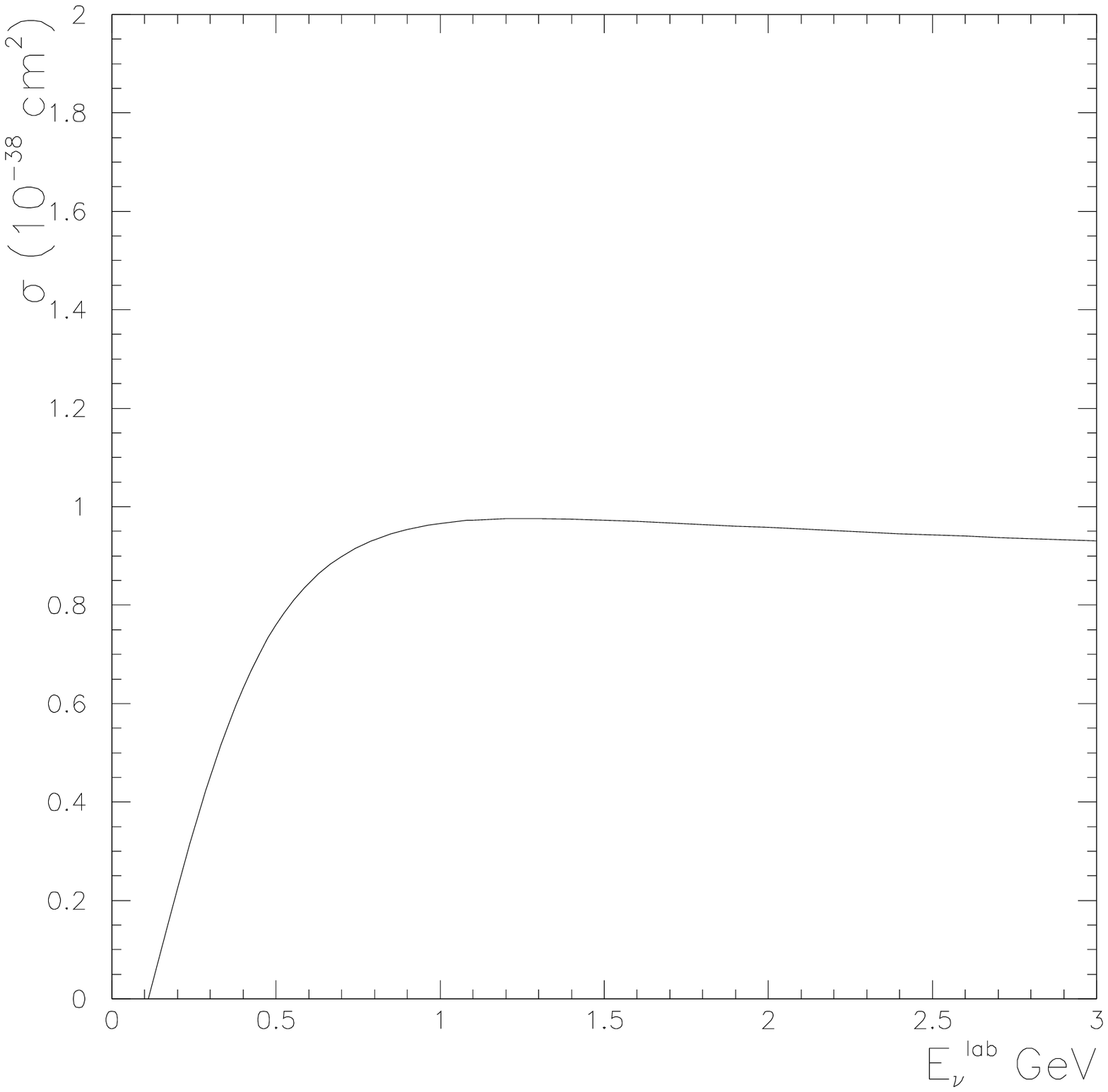,width=12.cm}
\end{center}
\caption{}
\label{fig:fig1}
\end{figure*}

In Fig.~\ref{fig:fig1} the cross section of quasi-elastic
reaction (\ref{eq:MM1}) as a function of the
neutrino energy in laboratory system is shown.
The cross section is calculated in the conventional theory with parameters and
form factors taken from Refs. \cite{Zeller,Llewellyn}. 
The strong dependence of the form
factors from a transfer momentum squared allows one to describe the
bechavior of this cross section of reaction (\ref{eq:MM1}) in a broad
energy range from
threshold and up to $E^{lab}_{\nu} \sim 10~GeV$. Nevertheless,
two different mechanisms of quasi-elastic reaction are
seen from Fig.~\ref{fig:fig1}. First is at energies
from a threshold of reaction and up to
$E^{lab}_{\nu} \sim 1.3~GeV$, and second is at $E^{lab}_{\nu}> 1.3~GeV$, i.e.
in the Bjorken scaling region.

To find the gauge-invariant amplitude of radiation process
(\ref{eq:MM5}) within the standard formalism, it is needed to consider
the diagrams of an internal Bremsstrahlung as well as to account for
the contributions of the direct vector and axial radiation. In any case  there
are
some free parameters, which have to be determined from the experiment. We are
interested in an estimation of the cross section of process (\ref{eq:MM5}) at
low neutrino energy $E^{lab}_{\nu}= 0.7~GeV$. The cross section of reaction
(\ref{eq:MM1}) at $E^{lab}_{\nu}= 0.7~GeV$ from our calculation
(see, Fig.~\ref{fig:fig1}) is:
\begin{eqnarray}
\label{eq:MM10}
\sigma \simeq 0.95 \times 10^{-38}~cm^2.  
\end{eqnarray}

In Ref. \cite{Efrosinin} for description of a hadron current in
$K_{\mu 3}$ decay the representation of current of constituent quarks was
used.

\begin{figure*}[hb]
\begin{center}
\epsfig{file=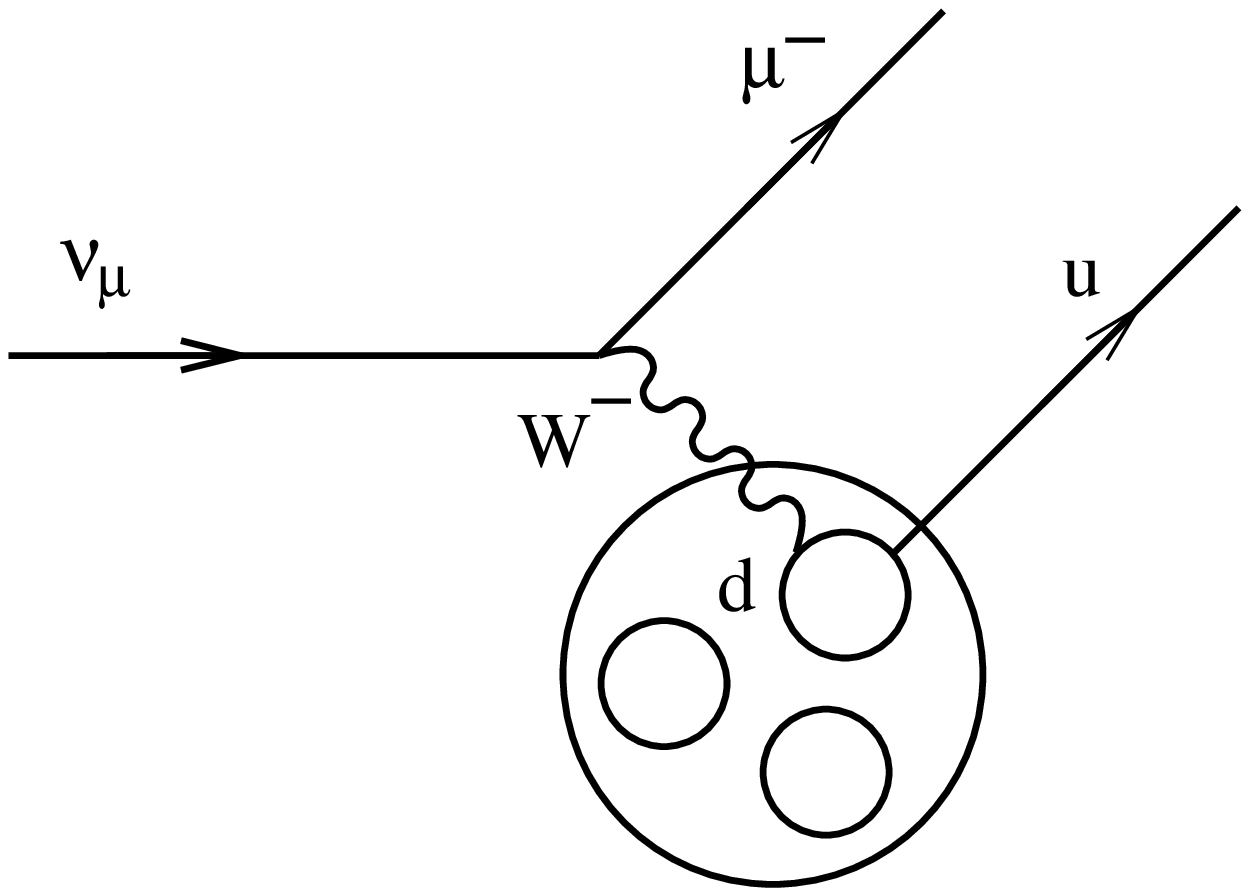,width=8.cm}
\end{center}
\caption{}
\label{fig:fig2}
\end{figure*}

We use this approach for the description of the amplitude of process
(\ref{eq:MM1}). We suppose that at the first stage of reaction
(\ref{eq:MM1}) neutrino interacts with $d$ quark, existing in the neutron,
with formation of muon and
$u$ quark with a definit momentum (see, Fig.~\ref{fig:fig2}). The produced
$u$ quark is going some time freely. At the second stage the confinement in a
proton demands that all system of three quarks is moving with an
produced $u$ quark momentum. The cross section of reaction
(\ref{eq:MM1}) is determined basically by the first stage of the process.

\begin{figure*}[hb]
\begin{center}
\epsfig{file=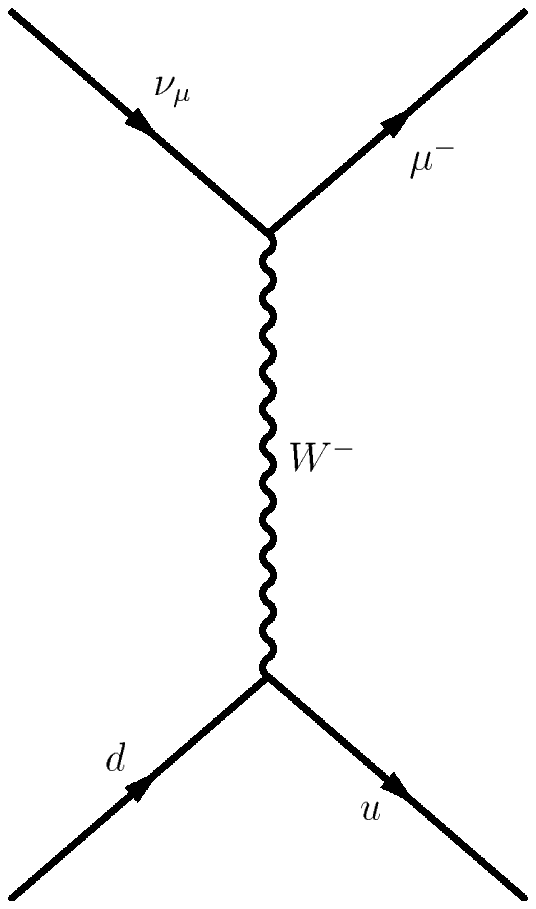,width=8.cm}
\end{center}
\caption{}
\label{fig:fig3}
\end{figure*}

\begin{figure*}[hb]
\begin{center}
\epsfig{file=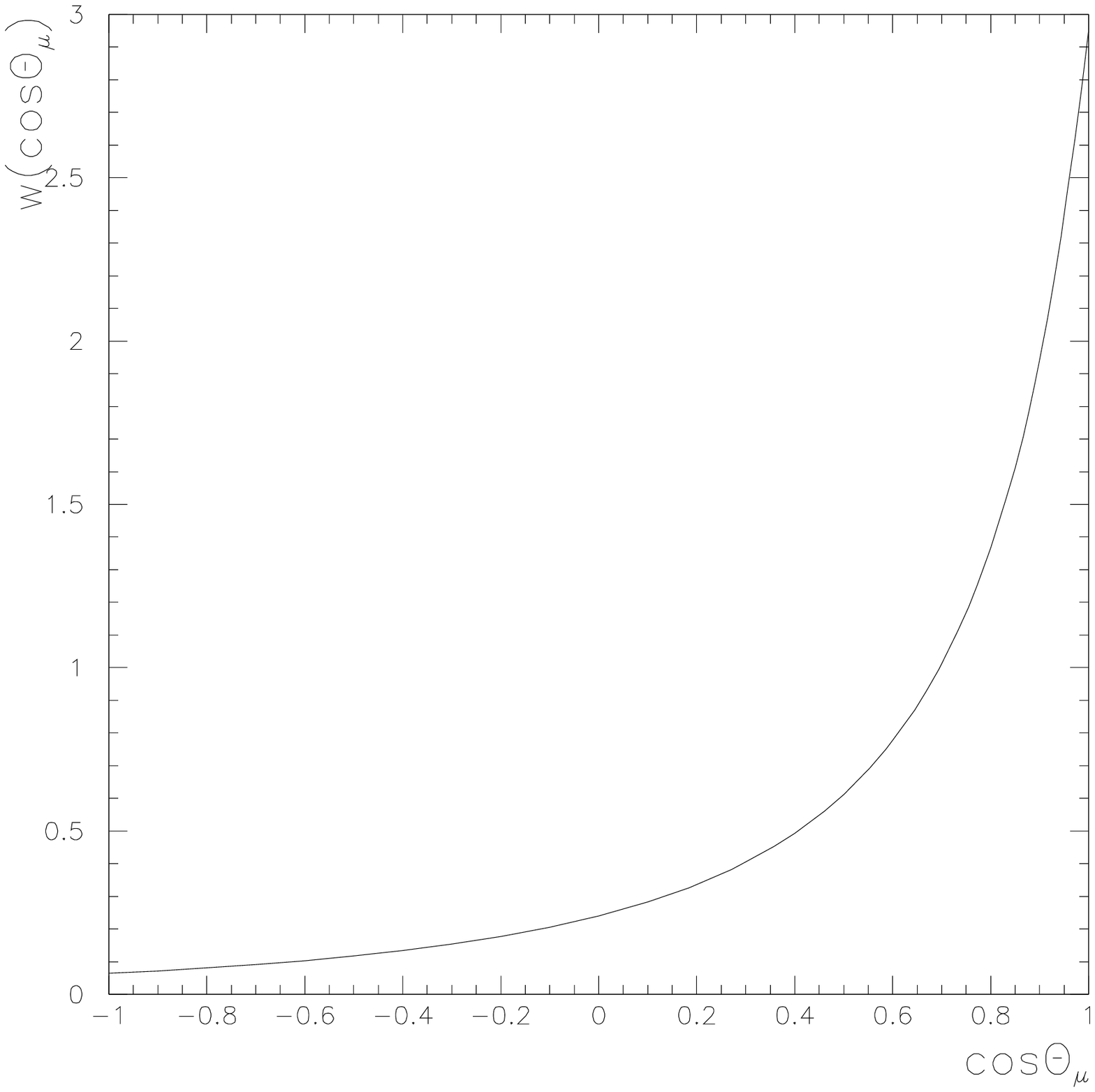,width=14.cm}
\end{center}
\caption{}
\label{fig:fig4}
\end{figure*}

The diagram of process (\ref{eq:MM1}) on quark level
\begin{eqnarray}
\label{eq:MM11}
\nu_{\mu}(k)+d(p) \to \mu^-(k^{\prime})+u(p^{\prime})  
\end{eqnarray}
is displayed in Fig.~\ref{fig:fig3}.
Amplitude of process (\ref{eq:MM1}) is given by
\begin{eqnarray}
\label{eq:MM12}
T=\frac{G_F}{\sqrt{2}}|V_{ud}|\bar{u}_u(p^{\prime})\gamma^{\mu}
(1-\gamma_5)u_d(p)\bar{u}_{\mu}(k^{\prime})\gamma_{\mu}
(1-\gamma_5)u_{\nu_{\mu}}(k).  
\end{eqnarray}
Here, $\bar{u}_u(p^{\prime}),u_d(p)$ are 4-spinors of constituent quarks,
the masses of constituent quarks as well as paper \cite{Efrosinin} and also in
papers \cite{Gershtein,Khlopov}, in particular values of masses $m_u=m_d=
0.305~GeV$ \cite{Efrosinin} were used. In Fig.~\ref{fig:fig4} the angular
distribution of a muon for reaction $\nu_{\mu}+n \to \mu^-+p$ in a laboratory
system at $E^{lab}_{\nu}= 0.7~GeV$ is shown. The quark amplitude
(\ref{eq:MM12}) determines an angular distribution of a muon in reaction
(\ref{eq:MM1}). The cross section of process (\ref{eq:MM11}) on one quark is
determined also imploing the amplitude (\ref{eq:MM12}) and at
$E^{lab}_{\nu}= 0.7~GeV$ it equals to
\begin{eqnarray}
\label{eq:MM13}
\sigma \simeq 0.53 \times 10^{-38}~cm^2.  
\end{eqnarray}
It is known that the point weak interaction is incoherent.
For obtaining the cross section of
reaction (\ref{eq:MM1}) on a neutron, containing two $d$ quarks, it is
necessary
to double the cross section (\ref{eq:MM13}). So, the cross section of
quasi-elastic
reaction $\nu_{\mu}+n \to \mu^-+p$ at $E^{lab}_{\nu}= 0.7~GeV$ is equal to
\begin{eqnarray}
\label{eq:MM13}
\sigma \simeq 1.06 \times 10^{-38}~cm^2.  
\end{eqnarray}
It corresponds to value (\ref{eq:MM10}) by an order of magnitude.

\begin{figure*}[hb]
\begin{center}
\epsfig{file=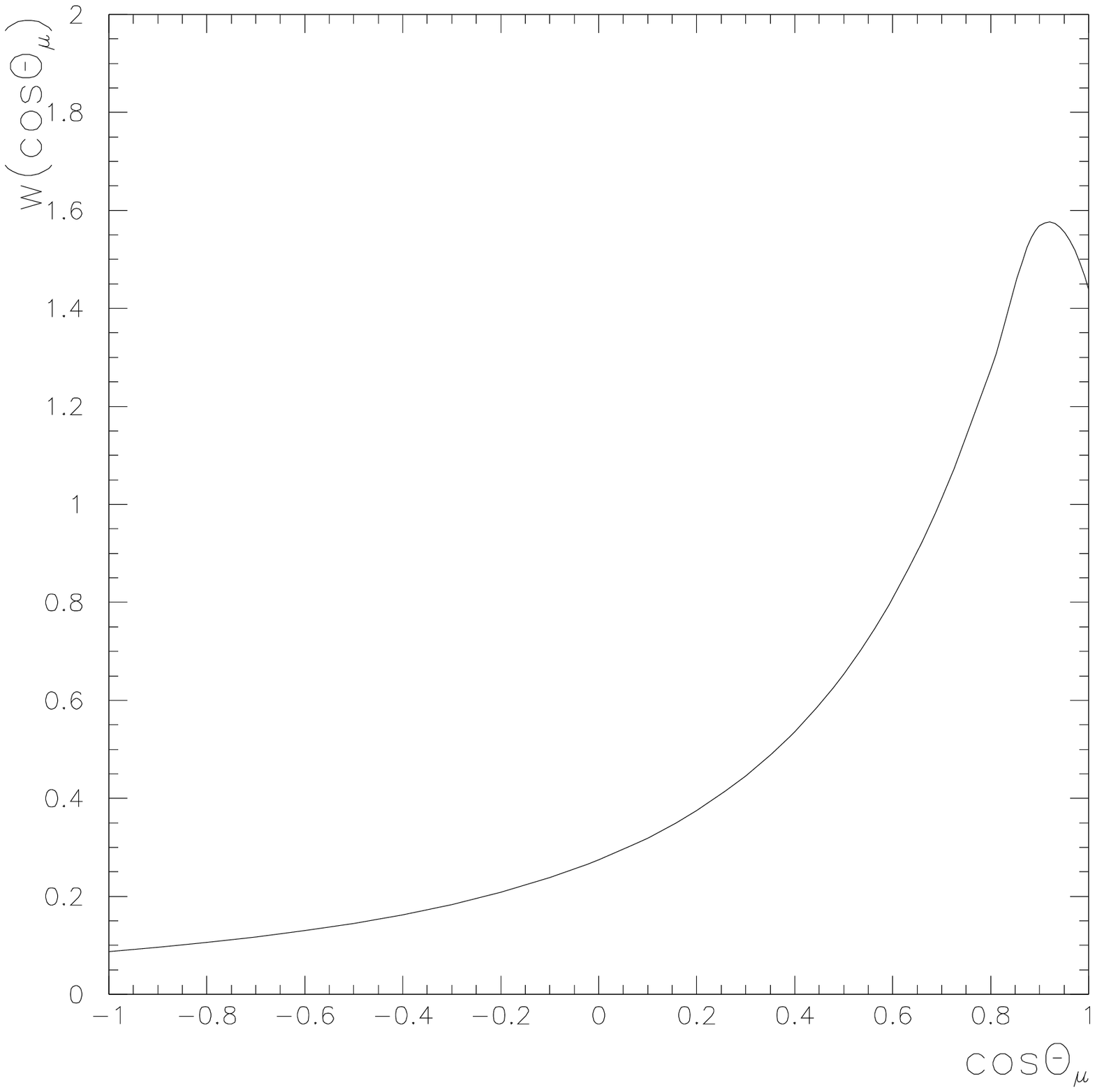,width=14.cm}
\end{center}
\caption{}
\label{fig:fig5}
\end{figure*}

In Fig.~\ref{fig:fig5} the similar angular distribution of muon in reaction
(\ref{eq:MM1}) in a laboratory sistem calculated using the conventional model
\cite{Zeller,Llewellyn} is shown. Let us note that
in K2K experiment on muon neutrino oscillation \cite{Aliu} the
deficit of
one-track events, whose muons are at angles near the direction of the neutrino
beam, was observed,
when Monte Carlo simulation with the conventional description of
quasi-elastic reaction was carried out. In an angular distribution of a muon,
shown in
Fig.~\ref{fig:fig4}, such
deficit of one-track events at small angles, contrary to the case of
Fig.~\ref{fig:fig5}, was not observed.

The conducted analysis allows one to make some qualitative conclusions:

1. The description of an experimental angular distribution of a muon in a
laboratory system in quasi-elastic reaction $\nu_{\mu}+n \to \mu^-+p$
is reached without use of the phenomenological form factors
(see, Fig.~\ref{fig:fig4}). The neutrino interacts with a quark
via the process (\ref{eq:MM11}).

2. In reaction $\nu_{\mu}+n \to \mu^-+p+\gamma$ the first stage of process
proceeds through the
interaction a neutrino with a simple system, a quark. Therefore, the direct
production of a photon does not take place. So, in general, the contributions
of direct radiation in the given reaction will be reduced.

\begin{figure*}[hb]
\begin{center}
\epsfig{file=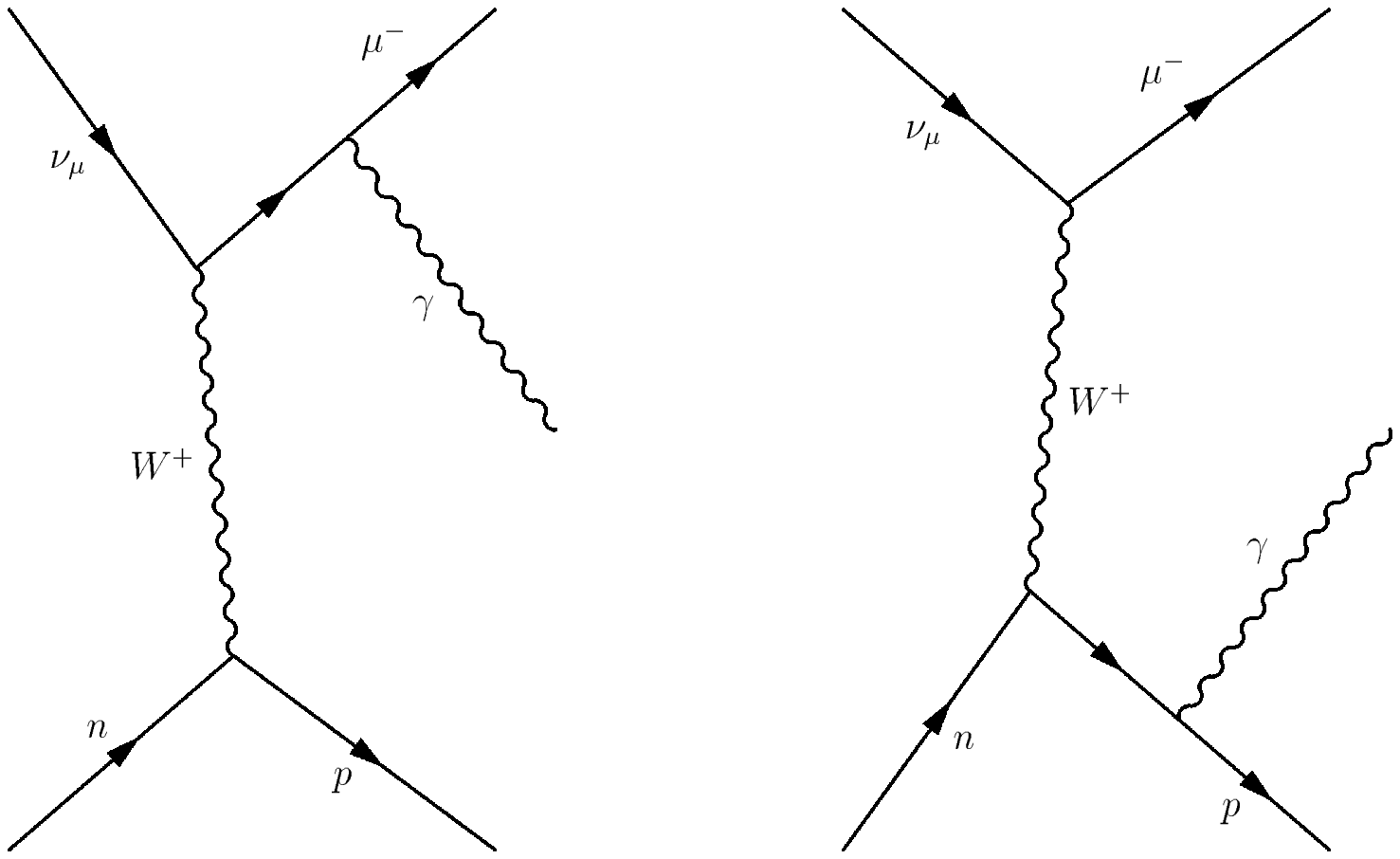,width=14.cm}
\end{center}
\caption{}
\label{fig:fig6}
\end{figure*}

Finally, we estimate the cross section of reaction
$\nu_{\mu}+n \to \mu^-+p+\gamma$. We can find the ratio of the cross
section of reaction (Fig.~\ref{fig:fig6}) 
\begin{eqnarray}
\label{eq:MM15}
\nu_{\mu}(k)+n(p) \to \mu^-(k^{\prime})+\gamma(r)  
\end{eqnarray}
and the cross section of reaction (\ref{eq:MM1}).

\begin{figure*}[hb]
\begin{center}
\epsfig{file=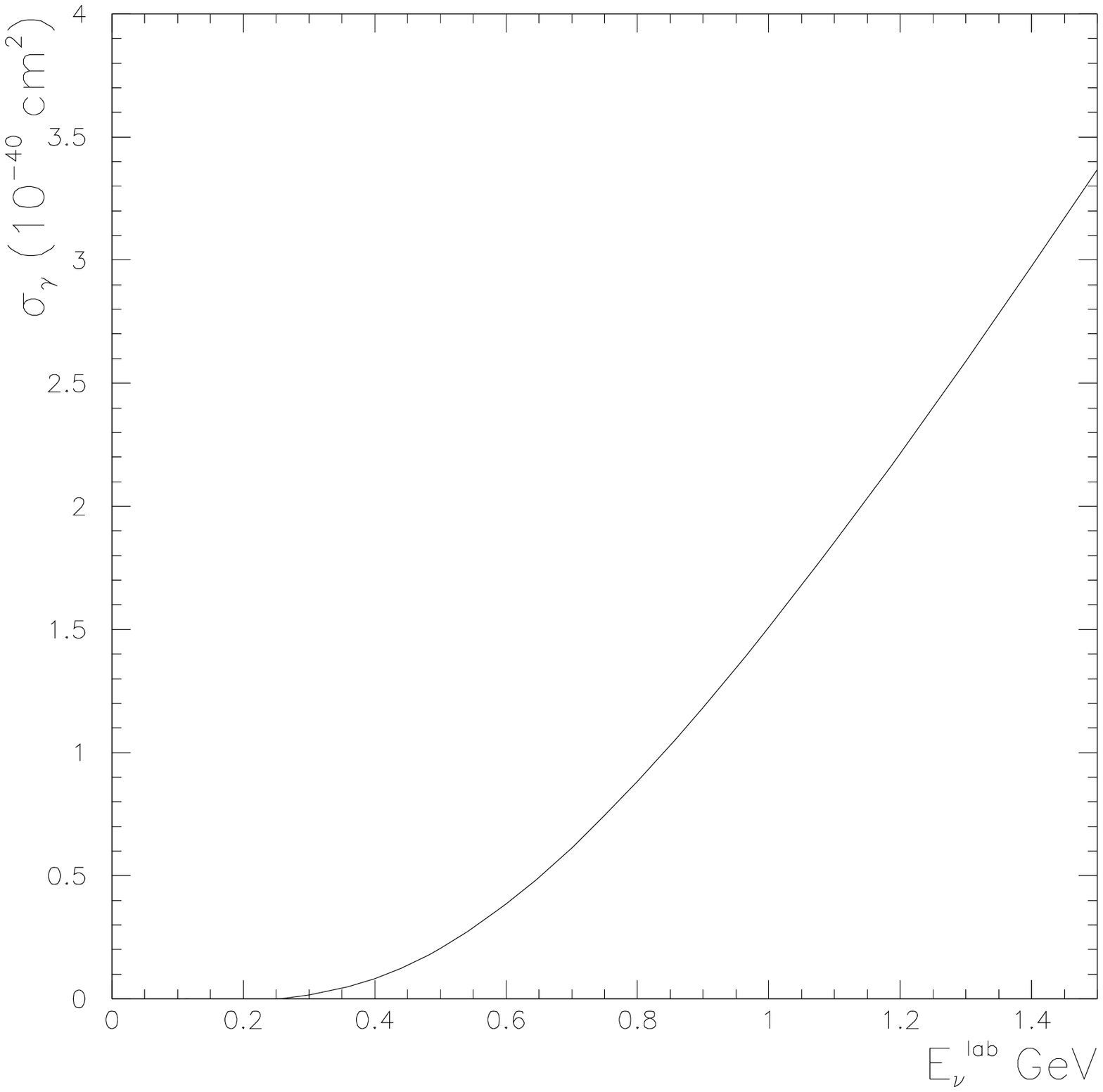,width=14.cm}
\end{center}
\caption{}
\label{fig:fig7}
\end{figure*}

\begin{figure*}[hb]
\begin{center}
\epsfig{file=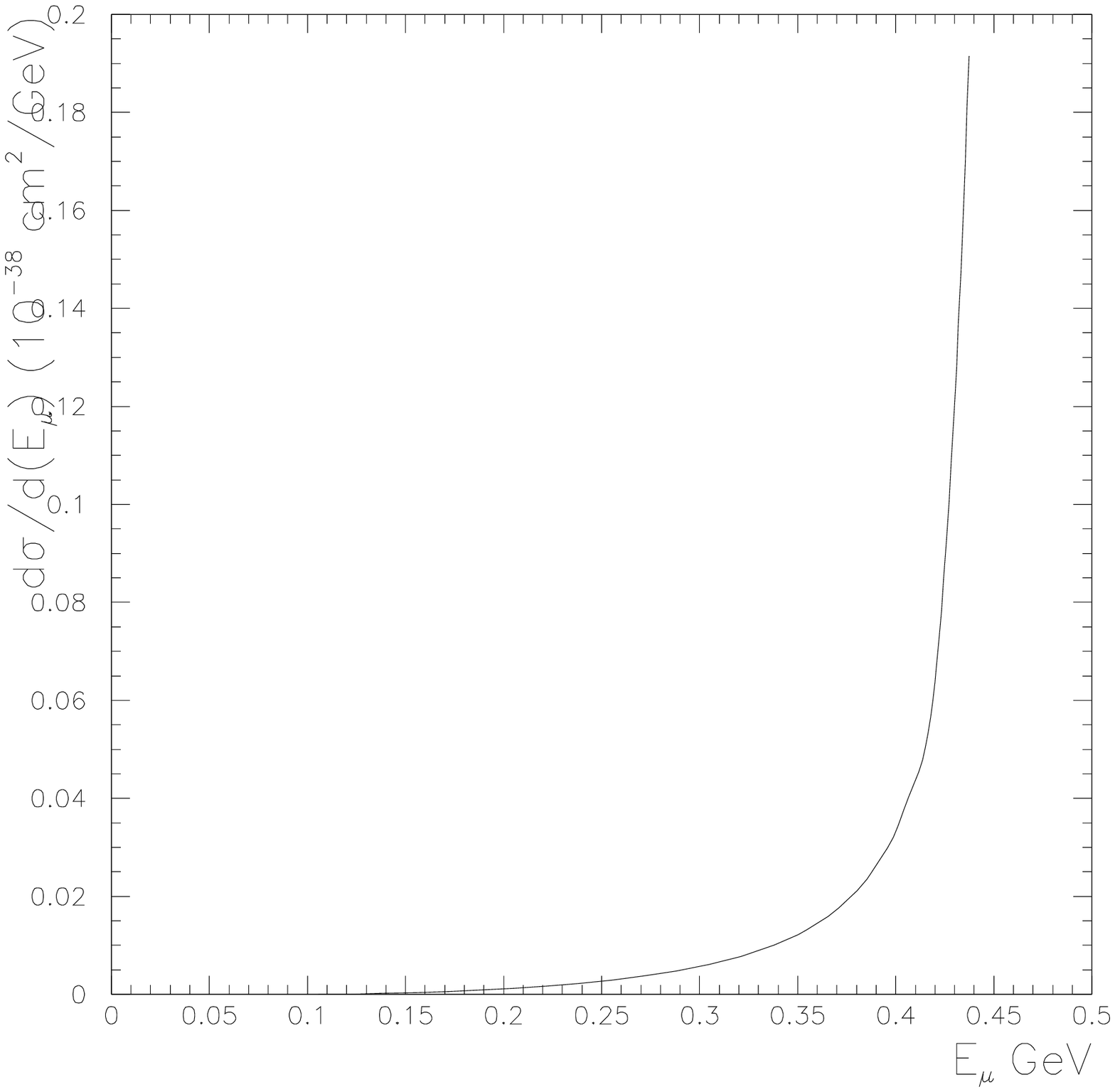,width=14.cm}
\end{center}
\caption{}
\label{fig:fig8}
\end{figure*}

\begin{figure*}[hb]
\begin{center}
\epsfig{file=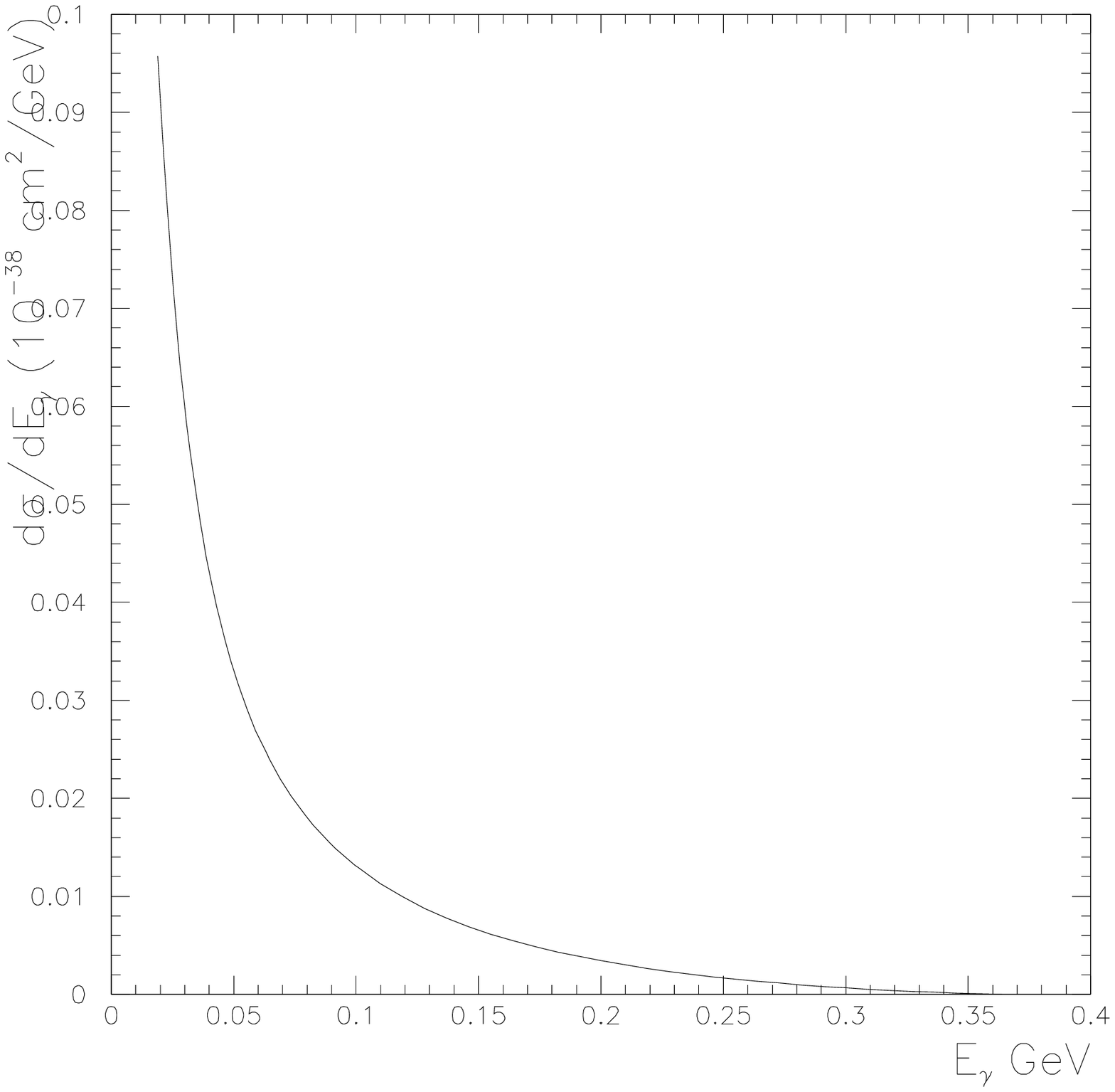,width=14.cm}
\end{center}
\caption{}
\label{fig:fig9}
\end{figure*}

\begin{figure*}[hb]
\begin{center}
\epsfig{file=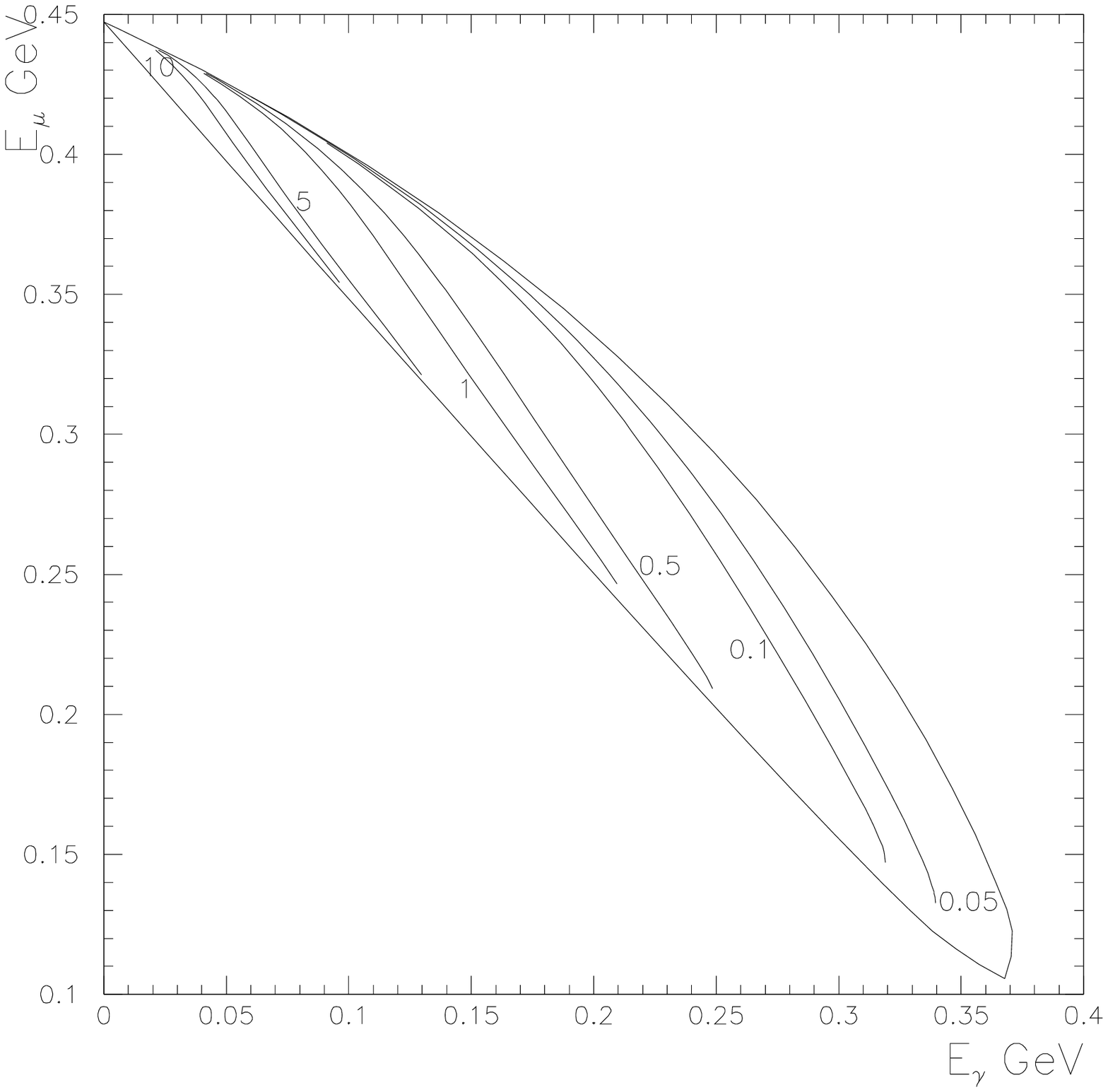,width=14.cm}
\end{center}
\caption{}
\label{fig:fig10}
\end{figure*}

Using the Low theorem \cite{low,chew,pest,baen,fisch}, the
gauge-invariant amplitude of reaction (\ref{eq:MM15}) can be written as
(within $r^1$):
\begin{eqnarray}
\label{eq:MM16}
M&=&e\frac{G_F}{\sqrt{2}}|V_{ud}|\Bigl[\frac{(k^{\prime}\varepsilon^*)}
{(k^{\prime}r)}-\frac{(p^{\prime}\varepsilon^*)}{(p^{\prime}r)}\Bigr]
\nonumber\\
&&\times\bar{u}(\tilde{p}^{\prime})\Bigl[F^1_V\gamma^{\mu}+F^2_V
\frac{i\sigma^{\mu\rho}q_{\rho}}{2m_N}+F_A\gamma^{\mu}\gamma_5
+F_p\frac{q^{\mu}\gamma_5}{m_N}\Bigr]u(p)\nonumber\\
&&\times\bar{u}(\tilde{k}^{\prime})\gamma^{\mu}(1-\gamma_5)u(k)+O(r).  
\end{eqnarray}
Here, $\varepsilon^*_{\lambda}$ is photon polarization, $\tilde{k}^{\prime}$
and $\tilde{p}^{\prime}$ denote the four-momenta of muon and proton in
quasi-elastic reaction (\ref{eq:MM1}); $q=k-\tilde{k}^{\prime}; F^1_V, F^2_V,
F_A$ are form factors of the conventional theory taken from Refs.
\cite{Zeller,Llewellyn}. We are suspected that the contributions of direct
radiation are insignificant at low neutrino energy. As a result of calculation,
the
value of the cross section of reaction (\ref{eq:MM15}) $\sigma_{\gamma}$ is
\begin{eqnarray}
\label{eq:MM17}
\sigma_{\gamma}=0.61 \times 10^{-40}~cm^2,  
\end{eqnarray}
at $E_{\gamma} \geq 0.01~GeV,
E^{max}_{\mu}-E_{\mu} \geq 0.01~GeV$ in center mass system.
It constitutes $0.65\%$ from the value of the cross section
$\sigma$ (\ref{eq:MM10}). The cross section for reaction
(\ref{eq:MM15}) as a function of laboratory neutrino energy is shown in
Fig.~\ref{fig:fig7}. In Fig.~\ref{fig:fig8} and Fig.~\ref{fig:fig9}
the energy distributions of scattered muon and photon for this reaction
at laboratory neutrino energy $E^{lab}_{\nu}= 0.7~GeV$ are shown. Dalitz plot
density for
reaction (\ref{eq:MM15}) at this energy $E^{lab}_{\nu}= 0.7~GeV$ is shown in
Fig.~\ref{fig:fig10}. In the adopted approximation
the angular distribution of a muon
in reaction (\ref{eq:MM15}) will be like on an angular distribution of a muon
in
quasi-elastic reaction (\ref{eq:MM1}). In this case angular distribution
of a photon in a center-of-mass system relatively to a direction of neutrino beam
will be flat.

The current experiments with high statistics on measurement of neutrino
oscillations will essentially influence on refinement of measurements of
cross sections of quasi-elastic reaction and attendant reactions. They will help
to better understand 
mechanism of interactions a neutrino of low energies with
nucleons.





\newpage
\clearpage

\newpage
\clearpage

\newpage
\clearpage

\newpage
\clearpage

\newpage
\clearpage

\newpage
\clearpage

\newpage
\clearpage

\newpage
\clearpage

\newpage
\clearpage

\newpage
\clearpage

\newpage
\clearpage

\begin{center}
Figure captions
\end{center}

Fig.~1. Cross sections for the quasi-elastic process
$\nu_{\mu}+n \to \mu^-+p$ as a function of neutrino energy in conventional
theory with parameters taken from \cite{Zeller,Llewellyn}.
		
Fig.~2. The neutrino interaction with $d$-quark in a laboratory system.			
		
Fig.~3. Diagram of process $\nu_{\mu}+n \to \mu^-+p$ on quark level at low
neutrino energies. 

Fig.~4. Angular distribution of a muon for reaction $\nu_{\mu}+n \to \mu^-+p$
in laboratory system at $E^{lab}_{\nu}= 0.7~GeV$ from quark model.

Fig.~5. Angular distribution of a muon for reaction $\nu_{\mu}+n \to \mu^-+p$
in laboratory system at $E^{lab}_{\nu}= 0.7~GeV$ from conventional theory.

Fig.~6. Diagrams of process $\nu_{\mu}+n \to \mu^-+p+\gamma$ on quark level
at low neutrino energies. 

Fig.~7. Cross section for reaction $\nu_{\mu}+n \to \mu^-+p+\gamma$ as
a function of laboratory neutrino energy.

Fig.~8. Energy distribution of scattered muons for reaction
$\nu_{\mu}+n \to \mu^-+p+\gamma$ at $E^{lab}_{\nu}= 0.7~GeV$

Fig.~9. Energy distribution of photon for reaction
$\nu_{\mu}+n \to \mu^-+p+\gamma$ at $E^{lab}_{\nu}= 0.7~GeV$.

Fig.~10. Dalitz plot density for reaction
$\nu_{\mu}+n \to \mu^-+p+\gamma$ at $E^{lab}_{\nu}= 0.7~GeV$.

.


\end{document}